\documentstyle[editedvolume,numreferences]{crckapb}

\begin{opening}
\title{SPONTANEOUS COALITION FORMING:\protect\\
       A MODEL FROM SPIN GLASS}
\subtitle{Eastern Europe post cold war instabilities}

\author{S. GALAM}
\institute{Laboratoire des Milieux D\'{e}sordonn\'{e}s et 
H\'{e}t\'{e}rog\`{e}nes,\\ Universit\'e Pierre et Marie Curie,
\\
Tour 13 - Case 86, 4 place Jussieu, 75252 Paris Cedex 05,
 \\ E-mail: galam@ccr.jussieu.fr}

\end{opening}

\runningtitle{SPONTANEOUS COALITION FORMING}

\begin{document}


\begin{abstract}
In this paper a simple model is proposed to decribe the spontaneous formation 
of coalitions among a group of actors like countries. The basic ingredients are 
from the physics of disorder systems.
It is the interplay of two different spin glass models with respectively
random bond
and random site disorders which is instrumental in the present
approach. The cold war stabilty period is then given
an explanation as well the instabilities produced by the Warsow pact dissolution.
European and Chiese stabilities are also discussed.

\end{abstract}
\section{Introduction}

In recent years, physics has been successful in providing
several models to describe some collective behaviors
in both human societies and social organisations \cite{1}.
In particular, new light has been shed on democratic voting biases \cite{2},
decision making process \cite{3}, outbreak of cooperation \cite{4}, 
social impact \cite{5},
and power genesis in groups \cite{6}.

However such a new approach to social behavior is yet
at its earlier stage. More work is needed as well more connexion
with social data. At this stage, it is really the first ingrdients to what 
could become, in the near future, a new field of research by itself. 
At least it is our challenge.

It is also worth to state few words of caution, since dealing with social
reality can often interfere with the reality itsel via biases in actual
social representations. One contribution of this ``sociophysics" would be 
indeed to
take away social studies from political or philosophical beliefs,
to place it on a more modelling frame without any religion like attitude.

In this paper we adress the question of coalition forming in the framework
of country military alliances, using some
basic concepts from the physics of spin glasses systems \cite{7}. 
Along this line, an earlier attempt from political sciences \cite{8},
used the physical concept of minimum energy. However 
this work was misleading \cite{9} since it was based on a confusion between 
the two physically different spin glass models
of Mattis and  Edwards-Anderson \cite{10}.
The model presented here, indeed involves the interplay between these two models.

The following of the paper is organised as follows. The second part
contains the presentation of our model. Several features in the dynamics 
of bimodal coalitions are obtained. Within such a framework the one country 
viewpoint is studied in Section 3 showing up the frontiers of turning some 
local cooperation to 
conflict or the opposite still preserving the belonging to the same coalition.
The setting up of world wide alliances is discussed in section 4. The cold war situation 
is then analysed
in section 5.  A new explanation is given in Section 6 to 
estern Europe instabilities following the Warsaw pact 
dissolution as well as to 
western Europe stability. Some hints are also obtained within
the model on how to stabilize these eastern Europe instabilities giving the 
still existing Nato
organisation. The model is then applied in Section 7 to describe
the Chinese situation. The concept of ``risky actor" is briefly introduced in Section 8.
Last Section contains some concluding remarks.

\section{Presentation of the model}

We now address the problem of alignment between a group of $N$ countries 
\cite{7}.
From historical, cultural and economic frames there exit bilateral
propensities $J_{i,j}$
between any pair
of countries $i$ and $j$ to either cooperation $(J_{i,j}>0)$, conflict
$(J_{i,j}<0)$
or ignorance $(J_{i,j}=0)$. Each propensity $J_{i,j}$ depends solely on the
pair $(i,\:j)$ itself
and is positive, negative or zero. Propensities $J_{i,j}$ are somehow local 
since they don't account for any global organization
or net.
Their intensities vary for each pair of countries to account
for the various military and economic power of both pair actors. 
They are assumed to be symmetric, i.e., $J_{ij}=J_{ji}$.

From the well known saying ``the enemy of 
an enemy is a friend" we postulate the existence of only  
two
competing coalitions, like for instance
western and eastern blocks during the so-called cold war.
They are denoted respectively by A
and B. 

Each actor has then the choice to be in either one of two coalitions. 
A variable $\eta _i$ associated to each actor,
where index i runs from 1 to N, states its actual belonging.
It is $\eta _i=+1$ if actor $i$ belongs to alliance A
while $\eta _i=-1$ in case it is part of alliance B. From symmetry all
A-members can
turn to coalition B with a simultaneous flip of all B-members to coalition
A.

Given a pair of actors $(i,j)$ their respective alignment is readily
expressed through
the product $\eta _i\eta _j$. The product is $+1$ when $i$ and $j$ belong
to the same coalition
and $-1$ otherwise. 
The ``cost" of exchange between
a pair of countries is then measured by the quantity $J_{ij}\eta _i\eta _j$.

Here factorisation over $i$ and $j$ is not possible. Indeed we are dealing
with competing given
bonds or links. It is equivalent to random
bond spin glasses as opposed to Mattis random site spin glasses \cite{10}.

Given a configuration $X$ of actors, for each nation $i$ we can measure
the overall degree of conflict and cooperation with all others $N-1$ countries,
with the quantity,
\begin {equation}
E_i=\sum^{n}_{j=1}J_{ij}\eta _j\,,
\end {equation}
where the summation is taken over all other
countries including
$i$ itself with $J_{ii}\equiv 0$. The product $\eta _iE_i$ then evaluates
the local ``cost" associated with country $i$ choice. It is positive if
$i$ goes along the tendancy produced by $E_i$ and negative otherwise.
For a given configuration $X$, all country local ``cost" sum up to
a total  ``cost",
\begin {equation}
E(X)=\frac{1}{2}\sum_i \eta _iE_i\,,
\end {equation}
where the $1/2$ accounts for the double counting of pairs.
This ``cost" measures indeed the level of satisfactions of each country
alliance choice. It can be recast as,
\begin{equation}
E(X)=\frac{1}{2}\sum_{i,j}^{n}J_{ij}\eta _{i}\eta _{j}\,,
\end{equation}
where the sum runs over the $n(n-1)$  pairs $(i,j)$. Eq.(3) is indeed
the Hamiltonian of an Ising random bond magnetic system.

\subsection{THE CHOSEN DYNAMICS}

At this stage we postulate that the actual configuration is the one which 
minimizes
the cost in each countiry choice. In order to favor two cooperating
countries $(G_{i,j}>0)$ in the same alliance,
we put a minus sign in from of the expression of Eq. (3), to get,
\begin{equation}
H=-\frac{1}{2}\sum_{i>j}^nJ_{ij}\eta _i\eta _j\,.
\end{equation}
There exist by symmetry $2^{n}/ 2$ distinct sets of
alliances since each country has 2 choices for coalition.

Starting from any initial configuration, the dynamics of the system is 
implemented by single actor coalition flips. An
actor turns to
the competing coalition
only if the flip decreases its local cost. The system has reached its
stable state once
no more flip occurs. Given $\{J_{ij}\}$, the $\{\eta _i\}$ are thus
obtained minimizing
Eq. (4). 

Since here the system stable configuration minimizes the
``energy", we are from the physical viewpoint, at the temperature $``T=0"$.
Otherwise when  $``T\neq 0"$ it is the
free-energy which has to be minimized. 
In practise for a finite system the theory can
tell which coalitions are possible and how many of them exist. But when
several coalitions
have the same energy, it is not possible to predict which one will be the
actual one.

\subsection{Frustration effect}

The physical concept of frustration \cite{10} is  embodied
in the model. For instance, in the case of three conflicting nations 
as Israel, Syria and Iraq, any possible alliance 
configuration leaves always someone unsatisfied.

To define this feture more precisely; let us attach
respectively
the labels 1, 2, 3 to each one of the three countries. In case we have equal
and negative exchange interactions
$J_{12}=J_{13}=J_{23}=-J$ with $J>0$, the associated minimum of the energy
(Eq. (4)) is equal to $-J$. However this value of the minimum is realized
for several possible
and equivalent coalitions. Namely for countries (1, 2, 3) we can have
respectively alignments
(A, B, A), (B, A, A), (A, A, B),
(B, A, B), (A, B, B), and
(B, B, A). First 3 are identical to last 3 by symmetry since here what
matters is which countries
are together within the same coalition. The peculiar property is
that the system never gets stable in just one configuration since it costs no
energy to switch from one onto another. This case is an archetype of
frustration.
It means in particular the existence of several ground states with exactly
the same energy.

Otherwise, for non equal interactions the system has one stable
minimum and no frustration occurs within the physical meaning defined above.
The fact that some interactions are not satisfied does not automatically
imply frustration in above sense of multiple equivalent set of alliances.

\section{A one country viewpoint}

We now make this point more quantitative within the present formalism.
Consider a given site $i$.
Interactions with all others sites can be
represented by a field,
\begin {equation}
h_i=\sum^{n}_{j=1}J_{ij}\eta _j\,
\end {equation}
resulting in an energy contribution
\begin {equation}
E_i=-\eta _ih_i\,,
\end {equation}
to the Hamiltonian $H=\frac{1}{2}\sum^{n}_{i=1}E_i$. Eq. (10) is minimum
for $\eta _i$ and $h_i$ having the same sign. For a given $h_i$ there
exists always
a well defined coalition except for $h_i=0$. In this case site $i$ is
``neutral" since then
both coalitions are identical with respect to its local ``energy" which
stays equal to zero.
A neutral site will flip with probability $\frac{1}{2}$. 

\subsection{SHIFTING COALITION}

The coupling $\{J_{ij}\}$ are given. Let us then assume there exists only 
one minimum.
Once the system reaches its stable equilibrium it gets trapped and the
energy is minimum.
At the minimum the field $h_i$ can be calculated for each site $i$ since
$\{J_{ij}\}$  are known as well as $\{\eta _{i}\}$.

First consider all sites which have the value -1. The existence of a unique
non-degenerate minimum makes associated fields  also negative.
We then take one of these sites, e.g. $k$, and
shift its value
from -1 to +1 by simultaneously changing the sign of all its interactions
$\{J_{kl}\}$ where
$l$ runs from 1 to $n$ ($J_{kk}=0$). This transformation gives,
\begin {equation}
\eta _{k}=+1\,{\rm and}\,h_k>0\,,
\end {equation}
instead of,
\begin {equation}
\eta _{k}=-1\,{\rm and}\,h_k<0\,,
\end {equation}
which means that actor $k$ has shifted from one coalition into the other one.

It is worth to emphazise that such systematic shift of propensities of
actor $k$ has no effect
on the others actors. Taking for instance actor $l$, its unique interaction
with actor $k$
is through $J_{kl}$
which did change sign in the transformation. However as actor $k$ has also
turn to the other
coalition, the associated contribution $J_{kl}\eta _{k}$ to field $h_l$ of
actor $l$ is
unchanged.

The shift process is then repeated for each member of actor k former coalition.
Once all shifts are completed there exits only one unique coalition.
Everyone is cooperating
with all others. The value of the energy minimum is unchanged in the
process.

Above transformation demonstrates the $\{J_{ij}\}$ determine the stable
configuration. It shows in particular that given any site configuration, it
always exists a set of
$\{J_{ij}\}$ which will give that configuration as the unique minimum of
the associated energy.
At this stage, what indeed matters are the propensity values. 

Above gauge transformation  shows what matters is the sign of
field
$\{h_{i}\}$ and not a given $J_{ij}$ value. A given set of field signs,
positive and negative,
may be realized through an extremely large
spectrum of $\{J_{ij}\}$.

This very fact opens a way to explore some possible deviations from a
national policy.
For instance given
the state of cooperation and conflict of a group of actors,
it is possible to find out limits in
which local pair propensities can be modified without inducing coalition shift.
Some country can turn from cooperation to conflict or the opposite, without
changing the belonging to a given alliance as long as the associated field
sign is unchanged. It
means that a given country could becomes hostile to some former allies,
still staying in the same
overall coalition. One illustration is given by german recognition of
Croatia against the will of
other european partners like France and England, without putting at stake
its belonging to the
European community. The Falklands war between England and Argentina is
another example since
both countries have strong american partnerships.

\section{Setting up coalitions}

From above analysis, countries were found to belong to some allianace without 
apriori macro-analysis at the regional or world level. Each country is
adjusting to its best interest withg respect to countries with whom
it interacts. However the setting up of global coalitions which are aimed 
to spread and organize economic and miulitaru exchanges  
produces an additional ingredient in each country choice.

Yet staying in the two coalition scheme, each country has an apriori
 natural choice.
To account for this fact we introduced for each actor $i$,
a variable $\epsilon _i$. It is $\epsilon
_i=+1$ if actor
would like to be in $A$, $\epsilon _i=-1$ in $B$ and $\epsilon _i=0$ for no
apriori.
Such natural belonging is induced by cultural and political history. 

Moreover we measure exchanges produced by these coalitions trough a set
of additional
pairwise propensities $\{C_{i,j}\}$. They are
always positive since sharing resources, informations, weapons is
basically profitable.
Nevertheless a pair $(i,\:j)$ propensity to cooperation, conflict or
ignorance is
$A_{i,j}\equiv \epsilon _i \epsilon _j C_{i,j}$ which can be positive,
negative or zero.
Now we do have a Mattis random site spin glasses \cite{10}.

Including both local and macro exchanges result in an overall pair propensity
\begin{equation}
O_{i,j}\equiv J_{i,j} +\epsilon _i \epsilon _j C_{i,j}\:,
\end{equation}
between two countries $i$ and $j$ with always $J_{i,j}>0$.

An additional variable $\beta_i=\pm 1$ is introduced to account for
benefit from economic and military pressure attached to a given alignment.
It is still
$\beta _i=+1$ in favor of $A$, $\beta _i=-1$ for $B$ and $\beta _i=0$ for
no belonging.
The amplitude of this economical and military interest is measured by a local
positive field $b_i$ which also accounts for the country size and its importance.
At this stage, the sets $\{\epsilon _i\}$ and $\{\beta _i\}$ are independent.

Actual actor choices to cooperate or to conflict result from the given set
of above quantites.
The associated total cost is,
\begin{equation}
H=-\frac{1}{2}\sum_{i>j}^n\{J_{i,j} +\epsilon _i \epsilon _j C_{ij}\}\eta
_i\eta _j
-\sum_{i}^n \beta _ib_i\eta _i \,.
\end{equation}

\section{Cold war scenario}

The cold war scenario means that the two existing world level coalitions
generate much stonger
couplings than purely bilateral ones, i.e., $|J_{i,j}|<C_{i,j}$
since to belong
to a world level coalition produces
more advantages than purely local unproper relationship.
In others words local propensities were unactivated since overwhelmed
by the two block trend. The overall system was very stable.
We can thus take $J_{i,j}=0$.
Moreover each actor must belong to a coalition, i. e.,
$\epsilon _i\neq 0$ and $\beta _i\neq 0$.
In that situation local
propensities to cooperate or to
conflict between two interacting countries result from
their respective individual macro-level coalition belongings. the cold war
energy is,
\begin{equation}
H_{CW}=-\frac{1}{2}\sum_{i>j}^n\epsilon _i \epsilon _j J_{ij}\eta _i\eta _j
-\sum_{i}^n \beta _ib_i\eta _i \,.
\end{equation}

\subsection{COHERENT TENDENCIES}

We consider first the coherent tendency case in which cultural and
economical trends go
along the same coalition, i.e., $\beta _i=\epsilon _i$. Then from Eq. (15)
the minimum of
$H_{CW}$ is unique with all country propensities satisfied.
Each country chooses its coalition  according to its natural belonging,
i.e., $\eta _i=\epsilon _i$.
This result
is readily proven via the variable change $\tau \equiv \epsilon _i \eta _i$
which
turns the energy to,
\begin{equation}
H_{CW1}=-\frac{1}{2}\sum_{i>j}^n J_{ij}\tau _i\tau _j
-\sum_{i}^n b_i\tau _i \,,
\end{equation}
where $C_{i,j}>0$ are positive constants. Eq. (16) is a ferromagnetic
Ising  Hamiltonian in positive symmetry breaking fields $b_i$. Indeed it
has one unique minimum with all $\tau _i=+1$.

The remarkable result here is that the existence of two apriori world level
coalitions is identical
to the case of a unique coalition with every actor in it. It shed light on
the stability of the Cold
War situation where each actor satisfies its proper relationship.
Differences and conflicts appear
to be part of an overall cooperation within this scenario.
Both dynamics are exactly the same since what matters is the existence of a
well
defined stable configuration. However there exists a difference which is
not relevant at this
stage of the model since we assumed $J_{i,j}=0$. However in reality
$J_{i,j}\neq 0$ making the existence of two coalitions to produce a lower
``energy" than
a unique coalition since then, more $J_{i,j}$ can be satisfied.

It worth to notice that field terms $b_i\epsilon _i \eta _i$ account
for the difference in energy cost in breaking a pair proper relationship
for respectively
a large and a small country.
Consider for instance two countries $i$ and $j$ with $b_i=2b_j=2b_0$.
Associated pair energy is
\begin{equation}
H_{ij}\equiv -J_{ij}\epsilon _i \eta _i\epsilon _j \eta _j-2b_0\epsilon _i
\eta _i
-b_0\epsilon _j \eta _j\,.
\end{equation}
Conditions $\eta _i=\epsilon _i$ and $\eta _j=\epsilon _j$ give the
minimum energy,
\begin{equation}
H_{ij}^m=-J_{ij}-2b_0-b_0\,.
\end{equation}
>From Eq. (18) it is easily seen that in case $j$ breaks proper alignment
shifting to
$\eta _j=-\epsilon _j$ the cost in energy is $2J_{ij}+2b_0$. In parallel
when $i$ shifts
to $\eta _i=-\epsilon _i$ the cost is higher with $2J_{ij}+4b_0$. Therfore
the cost in energy
is lower for a breaking from proper alignment by the small country
($b_j=b_0$) than by
the large country ($b_j=2b_0$).
In the real world, it is clearly not the
same for instance for the US to be against Argentina than to Argentina to
be against the US.

\subsection{UNCOHERENT TENDENCIES}

We now consider the uncoherent tendency case in which cultural and
economical trends may go
along opposite coalitions, i.e., $\beta _i\neq \epsilon _i$. Using above
variable change
$\tau \equiv \epsilon _i \eta _i$, the Hamiltonian becomes,
\begin{equation}
H_{CW2}=-\frac{1}{2}\sum_{i>j}^n J_{ij}\tau _i\tau _j
-\sum_{i}^n \delta _i b_i\tau _i \,,
\end{equation}
where $\delta _i \equiv \beta _i \epsilon _i$ is given and equal to $\pm1$.
$H_{CW2}$ is formally identical to the ferromagnetic Ising Hamiltonian in
random fields $\pm b_i$.
However, here the fields are not random.

The local field term $\delta _i b_i\tau _i$ modifies the country field
$h_i$ in Eq. (9) to
$h_i+\delta _i b_i$ which now can happen to be zero.
This change is qualitative since now there exists the possibility to have
``neutrality", i.e.,
zero local effective field coupled to the individual choice. Switzerland
attitude during World
war II may result from such a situation.
Moreover countries which have opposite cultural and economical trends may
now follow their
economical interest against their cultural interest or vice versa.
Two qualitatively different situations may occur.
\begin{itemize}
\item Unbalanced economical power: in this case we have $\sum_{i}^n\delta_i
b_i \neq 0$.

The symmetry is now broken in favor of one of the coalition. But still
there exists only one minimum.

\item Balanced economical power: in this case we have $\sum_{i}^n\delta_i
b_i = 0$.

Symmetry is preserved and $H_{CW2}$ is identical to the ferromagnetic Ising
Hamiltonian
in random fields which has one unique minimum.
\end{itemize}

\section{Unique world leader}

Now we consider current world situation where the eastern block has
disappeared. However it
is worth to emphazise the western block is still active as before in this
model. Within
our notations,
denoting $A$ the western alignment,
we have still $\epsilon _i=+1$ for countries which had
 $\epsilon _i=+1$. On the opposite, countries which had
$\epsilon _i=-1$ now turned to either $\epsilon _i=+1$ or
to $\epsilon _i=0$.

Therefore above $J_{i,j}=0$ assumption based on inequality
$|J_{i,j}|<|\epsilon _i\epsilon _j|C_{i,j}$ no longer holds for each pair
of countries.
In particular propensity $p_{i,j}$ becomes equal to $J_{i,j}$ in respective
cases where
$\epsilon _i=0$, $\epsilon _j=0$ and $\epsilon _i=\epsilon _j=0$.

A new distribution of actors results from the collapse of one block. On the
one hand $A$
coalition countries still determine their actual choices according to
$C_{i,j}$. On the other hand
former $B$ coaltion countries are now found to determine their choices
according to
competing links $J_{i,j}$ which did not automatically agree with former
$C_{i,j}$.
This subset of countries has turned from a Mattis random site spin glasses
without frustration
into a random bond spin glasses with frustration. In others world the
former $B$ coalition
subset has jumped from one
stable minimum to a highly degenerated unstable landscape with many local
minima.
This property could be related to the fragmentation process where ethnic
minorities and states
shift rapidly allegiances back and forth while they were part of a stable
structure just
few years ago.

While the $B$ coalition world organization has disappeared, the $A$
coalition world organization
did not change and is still active. It makes $|J_{i,j}|<C_{i,j}$ still
valid for $A$ countries
with $\epsilon _i\epsilon _j=+1$.
Associated countries thus maintain a stable relationship and avoid a
fragmentation process. This result supports a posteriori arguments
against the dissolution of Nato once Warsaw Pact was disolved.

Above situation could also shed some light on the european debate. It would mean
european stability is a result in particular of the existence of european
structures
with economical reality. These structures produce associated propensities
$C_{i,j}$
much stronger than local competing propensities $J_{i,j}$ which are still
there.
In other words european stability would indeed result from
$C_{i,j}>|J_{i,j}|$  and not from either all $J_{i,j}>0$ or all $J_{i,j}=0$.
An eventual setback of the european construction ($\epsilon _i\epsilon
_jC_{i,j}=0$)
would then automatically yield a
fragmentation process with activation of ancestral bilateral oppositions.

In this model, once a unique economical as well as military world level
organisation exists, each country interest becomes
to be part of it. We thus have $\beta _i=+1$ for each actor. There may
be some exception like Cuba staying almost alone in former $B$ alignment,
but this case will not
be considered here.
Associated Hamiltonian for the $\epsilon_i =0$ subset actor is,
\begin{equation}
H_{UL}=-\frac{1}{2}\sum_{i>j}^n G_{ij}\eta _i\eta _j
-\sum_{i}^n b_i\eta _i \,,
\end{equation}
which is formally equivalent to a random bond Hamiltonian in a field. At
this stage
$\eta _i=+1$ means to be part
of $A$ coalition which is an international structure. On the opposite
$\eta _i=-1$ is to be in a non-existing $B$-coalition which really means to
be outside of $A$.

For small field with respect to interaction the system may still exhibit
physical-like frustration
depending on the various $J_{i,j}$. In this case the system has many minima
with the same energy.
Perpetual instabilities thus occur in a desperate search for an impossible
stability.
Actors will flip continuously from one local alliance to the other. The
dynamics
we are refering to is an individual flip each time it decreases the energy.
We also allow
a flip with probabilty $\frac{1}{2}$
when local energy is unchanged.

It is worth to point out that only strong local fields may
lift fragmentation by putting every actor in $A$-coalition. It can be achieved
through economical help like for instance in Ukrainia. Another way is
military $A$ enforcement
like for instance in former Yugoslavia.

Our results point out that current debate over integrating former eastern
countries within Nato
is indeed relevant to oppose current fragmentation processes. Moreover it
indicated that an
integration would suppress actual instabilities
by lifting frustration.

\section{The case of China}

China is an extremely huge country built up from several very large states.
These state
typical sizes are of the order or much larger than most other countries in
the world.
It is therefore interesting to analyse China stability within our model
since it represents a case of simultaneous
Cold war scenario and Unique world leader scenario.

There exists $n$ states which are all part of
a unique coalition which is the chinese central state. Then all
$\epsilon_i=+1$ but $\beta_i =\pm 1$
since some states keep economical and military interest in the ``union"
$(\beta_i =+1)$
while capitalistic advanced rich states contribute more than their share to
the ``union"
$(\beta_i =-1)$. Associated Hamiltonian is,
\begin{equation}
H=-\frac{1}{2}\sum_{i>j}^n\{J_{i,j} + J_{ij}\}\eta _i\eta _j
-\sum_{i}^n \beta _ib_i\eta _i \,,
\end{equation}
where $C_{i,j}>0$ and $G_{ij}$ is positive or negative depending on
each pair of state $(i,\:j)$.

At this point China is one unified country which means in particular that
$C_{i,j}>|G_{ij}|$
for all pair of states with negative $G_{ij}$. Therefore $\eta _i=+1$ for
each state.
Moreover it also implies $b_i<q_iC_{i,j}$
where $q_i$ is the number of
states state $i$ interacts with. Within this model, three possible scenari
can be oulined with respect to China stability.
\begin{enumerate}
\item China unity is preserved.

Rich states will go along their actual economic growth with the central
power turning to a
capitalistic oriented federative like structure. It means turning all
$\epsilon_i$ to $-1$
with then $\eta_i=\epsilon_i$. In parallel additional development of poor
states is required in order
to maintain condition  $C_{i,j}>|G_{ij}|$ where some $G_{ij}$ are negative.

\item Some rich states break unity.

Central power is unchanged with the same political and economical
orientation making heavier
limitations over rich state development. At some point the condition
$b_i>q_iC_{i,j}$ may be achieved for these states. These very states will
then get a lower
``energy"
breaking down from chinese unity. They will shift to $\eta _i=-1$ in their
alignment with
the rest of China which has $\eta _j=+1$.

\item China unity is lost with a fragmentation phenomenon.

In this case, opposition among various states becomes stronger than the
central organisational
cooperation with now $C_{i,j}<|G_{ij}|$ with some negative $G_{ij}$. The
situation would become
spin glass-like and
China would then undergo a fragmentation process. Former China would become
a highly unstable part of the world.
\end{enumerate}

\section{The risky actor driven dynamics}

In principle actors are expected to follow their proper relationship, i.e.,
to minimize their local ``energy". In other words, actors
follow normal and usual patterns of decision. But it is well known
that in real life these expectations are sometimes violated. Then
new situations are created with reversal of on going
policies.

To account for such situations we introduce the risky actor. It is an actor
who goes against his well defined interest. It
is different from the frustrated actor which does not have a well defined
interest.
Up to now everything was done at  $``T=0"$. However a risky actor chooses
coalition associated
to $\eta _i=-1$, although its local field
$h_i$ is positive. Therefore the existence of risky actors requires a
$T\neq 0$ situation.
The case of Rumania, having its own independent foreign policy, in former
Warsaw Pact may be an
illustration
of risky actor behavior. Greece and Turkey in the Cyprus conflict may be
another example.

Once $T\neq0$, it is not the energy which has to be minimized but the free
energy,
\begin{equation}
F=U-TS\,,
\end{equation}
where U is the internal energy, now different from the Hamiltonian and
equal to its thermal
average and S is the entropy. To minimize the free energy means stability
of a group of
countries matters on respective size of each coalition
but not, which actors are actually in these
coalitions. At a fixed "temperature" we thus can expect simultaneous shift
of alliances from several
countries as long as the size of the coalition is unchanged, without any
modification in the
relative strenghts. Egypt quitting soviet camp in the seventies and
Afghanistan joining it
may illustrate these non-destabilizing shifts.

Within the coalition frame temperature could be viewed as a way to account
for some risky trend.
It is not possible to know which particular actor will take a chance but
it is reasonable
to assume the existence of some number of risky actors. Temperature would
thus be a way to
account for some global level of risk taking.

Along ideas developped elsewhere \cite{6,11} we can assume that a level
of risky behavior is
profitable for the system as a whole. It produces surprises which induce to
reconsider some
aspect of coalitions themselves. Recent danish refusal to the signing of
Maastricht agreement
on closer european unity
may be viewed as an illustration of a risky actor. The net effect have been
indeed to turn what
seemed a trivial and apathetic administrative agreement into a deep and
passionated debate
among european countries with respect to european construction.

Above discussion shows implementation of $T\neq 0$ within the present
approach of coalition
should be rather fruitful.
More elaboration is left for future work.


\section{Conclusion}

In this paper we have proposed a new way to understand 
the alliance forming phenomena. in particular it was shown that within our 
model the cold war stabilty was not the result of two opposite alliances
but rather the existence of alliance induced exchange which neutralize 
the conflicting interactions within allies. It means that to have two  
alliances or jut one is qualitatively the same with respect to stability.

From this viewpoint the strong instabilies which resulted from
the Warsow pact dissolution  are given a simple explanation.
Simultaneously some hints are obtained about possible policies to
stabilize world
nation relationships. Along this line, the importance of european construction 
was also underlined.

We have also given some ground to introduce non-rational behavior in country
behaviors.
especially with the notions of "risky", "frustrated"
or ``neutral" actors.
A "risky" actor acts against his well defined interest while a "frustrated"
actor acts randomly since not having a
well defined interest. .

At this stage, our model remains rather primitive. However 
it opens some new road to explore and to forecast international policies.

\subsection*{Acknowledgments}
I indebted to D. Stauffer for numerous comments and critical discussions on
the manuscript.


\end{document}